\pdfoutput=1

\documentclass[11pt]{article}

\usepackage[preprint]{acl}

\usepackage{bm}

\usepackage{CJKutf8}
\usepackage{makecell}
\usepackage{graphicx}
\usepackage{amsmath}
\usepackage{amsfonts}
\usepackage{algorithm, algpseudocode}
\usepackage{hyperref}
\usepackage{multirow}
\usepackage{booktabs}
\usepackage{color}
\algrenewcommand\algorithmicrequire{\textbf{Input:}}
\algrenewcommand\algorithmicensure{\textbf{Output:}}

\usepackage{times}
\usepackage{latexsym}

\usepackage[T1]{fontenc}

\usepackage[utf8]{inputenc}

\usepackage{microtype}

\usepackage{inconsolata}

\usepackage{graphicx}

\title{Vulnerabilities of Audio-Based Biometric Authentication Systems Against Deepfake Speech Synthesis}

\author{Mengze Hong$^{1}$, Di Jiang$^{1}$\thanks{Corresponding Author}, Zeying Xie$^{2}$, Weiwei Zhao$^{2}$\\ \textbf{Guan Wang}$^{1}$, \textbf{Chen Jason Zhang}$^{1}$\\
$^{1}$Hong Kong Polytechnic University, $^{2}$AI Group, WeBank Co., Ltd
}

\begin{document}

\maketitle

\begin{abstract}

As audio deepfakes transition from research artifacts to widely available commercial tools, robust biometric authentication faces pressing security threats in high-stakes industries. This paper presents a systematic empirical evaluation of state-of-the-art speaker authentication systems based on a large-scale speech synthesis dataset, revealing two major security vulnerabilities: 1) modern voice cloning models trained on very small samples can easily bypass commercial speaker verification systems; and 2) anti-spoofing detectors struggle to generalize across different methods of audio synthesis, leading to a significant gap between in-domain performance and real-world robustness. These findings call for a reconsideration of security measures and stress the need for architectural innovations, adaptive defenses, and the transition towards multi-factor authentication.

\end{abstract}

\section{Introduction}
Voiceprint-based biometric authentication is a critical security modality, widely relied upon to safeguard financial transactions, verify identities remotely, control secure access, and prevent fraud across telecommunications systems \cite{li2024audio, kamel2025surveythreatsvoiceauthentication}. The global voiceprint authentication market is projected to grow from USD 2.87 billion in 2025 to USD 15.69 billion by 2032 \cite{alsheavi2025iot}. However, this reliance faces increasing vulnerabilities. Audio deepfakes have rapidly evolved from a laboratory curiosity into tangible real-world security threats \cite{rabhi2024audio}, causing significant societal and financial losses, including AI-generated robocalls impersonating public figures that reached millions of recipients and voice cloning scams defrauding the elderly of over \$200K \cite{alali2025partial, mittal2025pitch, shapiro2025cyber}. This crisis calls for the urgent assessment of the resilience of modern defense systems against deepfake technologies.

In this paper, we present rigorous empirical evaluations addressing the critical question of \textbf{whether state-of-the-art speaker authentication systems can withstand contemporary open-source voice cloning (deepfake) models}, which can synthesize speech from only a few minutes of target speaker data \cite{li2025survey}. We construct a large-scale benchmark by training multiple representative voice cloning systems on Mandarin speakers and evaluating both commercial speaker verification platforms and state-of-the-art anti-spoofing detectors across diverse settings to reveal their true security level and robustness. Our results reveal two previously undocumented security vulnerabilities: (1) speaker verification provides only partial defense against modern voice cloning attacks; and (2) anti-spoofing detectors fail to generalize effectively to unseen synthesis patterns. Both of these findings are amplified by the rapid evolution of speech synthesis technology and demand proactive attention from industry and academia. The contributions of this work are summarized as follows:

\begin{itemize}
    \item We present the first systematic evaluation of audio-based authentication systems under deepfake attacks, revealing critical vulnerabilities in state-of-the-art models.
    \item We identify and characterize key failure modes underlying system vulnerabilities, highlighting architectural limitations, the role of training data diversity in generalization, and the impact of self-supervised pretraining on cross-lingual transferability.
    \item We outline concrete directions for future research, emphasizing the need for continuously updated training corpora and architectures that capture intrinsic synthesis characteristics to enable a truly effective defense layer.
\end{itemize}

\section{Related Work}

\paragraph{Voice Cloning.} Modern voice cloning systems have advanced from requiring hours of target data to operating with just minutes of sample speech. Table~\ref{tab:attacks} summarizes mainstream systems, including text-to-speech (TTS), which generates speech from textual input, and voice conversion (VC), which transforms one speaker's voice to sound like another \cite{kaur2023conventional}. Open-source models such as GPT-SoVITS, Bert-VITS2, and RVC require only a few minutes of target speech and can be trained on a single V100 GPU within a few hours. Commercial models, by contrast, require even less training data at a reasonable cost, substantially lowering the barrier for malicious use relative to earlier ASVspoof-era attacks that required a large amount of data and computational resources \cite{todisco2019asvspoof}.

\begin{table}[t]
\centering
\small
\resizebox{\columnwidth}{!}{
\begin{tabular}{lccl}
\toprule
Model & Open-sourced & Data (mins) & Time / Cost \\
\midrule
\multicolumn{4}{c}{\textit{Text-to-Speech}} \\
\midrule
GPT-SoVITS  & \checkmark & 0.5 -- 2  & $\sim$10 min \\
Bert-VITS2  & \checkmark & 1 -- 5 & $\sim$2 h \\
ElevenLabs  & \texttimes & 2 -- 30 & $\sim\$0.73$ \\
Doubao & \texttimes & 0.5 -- 2 & $\sim\$15$ \\
Aliyun & \texttimes & 20 -- 30 & $\sim\$645$ \\
\midrule
\multicolumn{4}{c}{\textit{Voice Conversion}} \\
\midrule
RVC         & \checkmark & 10 -- 30 & $\sim$2 h \\
\bottomrule
\end{tabular}}
\caption{Comparison of modern speech synthesis systems by open-source availability, required target speaker data, and training time or cost per speaker.}\label{tab:attacks}
\end{table}

\paragraph{Audio Deepfake Detection.} Audio deepfake detection can be broadly divided into pipeline detectors, which combine hand-crafted features with classifiers, and end-to-end models that operate directly on raw waveforms \cite{li2025survey}. Pipeline approaches typically use LFCC, MFCC, or CQCC features \cite{todisco2016new, todisco2018integrated}, while end-to-end models exploit raw waveform representations \cite{tak2021end, hua2021towards}. Recent advances leverage self-supervised learning (SSL) and hybrid strategies to improve robustness: \citet{ge2025post} proposed post-training SSL models to enhance generalization to unseen attacks, while \citet{tahaoglu2025deepfake} proposed a ResNeXt-based architecture with spectral features to improve detection reliability.

\paragraph{Robustness and Generalization.} Real-world deployment requires authentication systems that are both robust and generalizable. Prior work has studied robustness to environmental factors such as codec compression, transmission noise, and reverberation \cite{tak2022rawboost}, as well as cross-dataset generalization, where models trained on one corpus often suffer significant performance drops on another \cite{wang2021comparative}. However, systematic evaluation across diverse synthesis architectures remains limited, representing a critical gap that fundamentally breaks system security.

\section{Experiment Setup}

To systematically evaluate authentication robustness against voice cloning attacks, we propose a framework integrating state-of-the-art speaker verification models, anti-spoofing detectors, and diverse speech synthesis approaches\footnote{Code and dataset will be released upon acceptance.}.

\subsection{Speaker Verification Model}
We employ the emerging ECAPA-TDNN architecture for speaker verification \cite{serre2025contrastive}. The model uses a time delay neural network (TDNN) backbone with channel-wise and context-wise attention mechanisms to extract discriminative speaker embeddings. We train the system on VoxCeleb \cite{nagrani2017voxceleb}, a large-scale dataset for speaker recognition with over one million utterances spanning 2,000+ hours. The detection threshold is tuned on the development set with a false acceptance rate of 0.01\%, following standard practice in compliance-critical applications \cite{brydinskyi2024comparison}.

\subsection{Deepfake Detection Model}

To detect deepfake speech, we adopt a state-of-the-art architecture combining XLS-R \cite{zhang2024audio}, a multilingual self-supervised speech representation model pretrained on 436k hours of multilingual speech, with AASIST \cite{zhang2024improving, jung2022aasist}, a graph attention-based spoofing detector. This system captures both rich semantic features and fine-grained spoofing artifacts, with proven performance on various benchmarks \cite{yamagishi2021asvspoof, tran2025multi}.

\begin{table}[t]
\centering
\small
\resizebox{\columnwidth}{!}{
\begin{tabular}{l l c c}
\toprule
& \textbf{Source} & \textbf{\# Speakers} & \textbf{Total Duration} \\
\midrule
Genuine & AISHELL-3 & 50 & 1000 \\
\midrule
\multirow{3}{*}{Synthetic} 
& GPT-SoVITS & 50 & 1000 \\
& Bert-VITS2 & 50 & 1000 \\
& RVC & 50 & 1000 \\
\bottomrule
\end{tabular}}
\caption{Overview of the benchmark dataset, with total duration in minutes.}\label{tab:dataset}
\end{table}

\subsection{Benchmark Dataset}
We randomly select 50 speakers (25 male, 25 female) from the AISHELL-3 dataset in Chinese Mandarin \cite{shi2020aishell}. For each speaker, 20 minutes of genuine speech are used to train three open-sourced voice synthesis systems\footnote{Project pages: \href{https://github.com/RVC-Boss/GPT-SoVITS}{GPT-SoVITS}, \href{https://github.com/fishaudio/Bert-VITS2}{Bert-VITS2}, and \href{https://github.com/RVC-Project/Retrieval-based-Voice-Conversion-WebUI}{RVC}.}: GPT-SoVITS and Bert-VITS2 for text-to-speech, and RVC for voice conversion. Each system generates 20 minutes of synthetic speech per speaker (see Table \ref{tab:dataset}). The dataset is split by speaker: 30 for training, 10 for development, and 10 for testing, ensuring that the test set remains entirely unseen during model training.

\subsection{Evaluation Metric}
In speaker verification, the bypass rate denotes the fraction of attacks that are misclassified as the target speaker. For deepfake detection, performance is evaluated using the Equal Error Rate (EER) \cite{reis2016audio}, defined as the point where the False Acceptance Rate equals the False Rejection Rate. The EER summarizes the trade-off between accepting spoofed speech as genuine and rejecting genuine speech as spoofed. Lower EER values indicate better discrimination, with 0\% representing perfect performance.

\section{Results and Discussions}

\begin{table}[!t]
\centering
\small
\begin{tabular}{lcc}
\toprule
\textbf{Synthesis Model} & \textbf{Bypass Rate} & \textbf{Avg. Similarity} \\
\midrule
GPT-SoVITS  & 56.2\% & 0.598 \\
Bert-VITS2  & 82.7\% & 0.679 \\
RVC         & 43.1\% & 0.558 \\
\bottomrule
\end{tabular}
\caption{Voiceprint verification bypass rates. }
\label{tab:voiceprint_detailed}
\end{table}

\subsection{Speaker Verification Vulnerability}

Table~\ref{tab:voiceprint_detailed} shows high bypass rates across all three voice cloning systems against the SOTA speaker verification model, revealing a key vulnerability: although the system achieves very low false acceptance on genuine users (FAR = 0.01\%), it fails to distinguish high-quality synthetic speech that closely mimics the speaker’s voiceprint characteristics. The average cosine similarity for all attacks exceeds 0.55, approaching the typical range of 0.6 -- 0.8 observed for genuine same-speaker utterances. Our findings reveal an alarming insight: the voiceprint authentication systems relied upon by millions of users worldwide can be easily compromised with just 10 -- 30 minutes of target speech, which is readily available from social media, podcasts, or public speeches. The barrier to attack is remarkably low, as a single consumer-grade GPU can train the required models in less than 2 hours. Together, these results expose a serious and actionable security risk, underscoring the urgent need for robust defenses against synthetic voice attacks.

\subsection{In-Domain Deepfake Detection}

\begin{table}[t]
\centering
\small
\resizebox{\columnwidth}{!}{
\begin{tabular}{lc}
\toprule
\textbf{Model} & \textbf{EER (\%)} \\
\midrule
LFCC + GMM \cite{todisco2018integrated}        & 12.43 \\
ResNet34 \cite{he2016deep} (spectrogram input)  & 3.21 \\
RawNet2 \cite{tak2021end}     & 2.14 \\
AASIST (standalone) \cite{jung2022aasist}      & 1.37 \\
\midrule
\textbf{XLS-R + AASIST}         & \textbf{0.83} \\
\bottomrule
\end{tabular}}
\caption{Comparison of deepfake detection models with in-domain test set.}
\label{tab:baselines}
\end{table}

On the in-domain test set, where both training and testing data are generated from the same group of deepfake models, XLS-R + AASIST achieves an EER of 0.83\%, significantly outperforming traditional methods and demonstrating its potential as a robust layer in an authentication system (see Table \ref{tab:baselines}). However, we argue that such performance does not reliably reflect practical robustness. In real-world scenarios, attackers can choose from a wide range of deepfake models, leading to out-of-domain conditions where attack speech is generated by models with synthesis patterns unseen during training. This necessitates further evaluations on robustness, a step often overlooked in existing studies that result in the misalignment between perceived and true robustness during deployment.

\subsection{Robustness Analysis}

\subsubsection{Generalization to Unseen Speech Synthesis Models}

To evaluate robustness against unseen attack models, we expand the test set with speech generated by \textbf{eight cutting-edge TTS systems} that differ from those used in training. The expanded test set contains 326 utterances (157 real, 169 synthetic), primarily sourced from closed-source or demo-only models\footnote{See Appendix \ref{appendix:model list} for the full list of models.}. These systems span diverse synthesis paradigms, including flow-based, diffusion-based, and prompt-conditioned architectures, enabling an assessment of whether the detector learns generalizable features rather than model-specific signatures.

\begin{table}[!t]
\centering
\small
\resizebox{\columnwidth}{!}{
\begin{tabular}{lcc}
\toprule
\textbf{Test Set} & \textbf{EER (\%)} & \textbf{Performance Gap} \\
\midrule
In-domain & 0.83 & - \\
\midrule
Out-of-domain & 24.84 & 29.9$\times$ \\
\midrule
ASVspoof 2021 LA & 3.48 & 4.2$\times$ \\
ASVspoof 2021 DF & 4.59 & 5.5$\times$ \\
\bottomrule
\end{tabular}}
\caption{Performance degradation of deepfake detection from in-domain to out-of-domain (model variation) and cross-language attacks.}
\label{tab:ood_detailed}
\end{table}

As shown in Table \ref{tab:ood_detailed}, a 30$\times$ performance degradation reveals a critical security risk: despite near-perfect in-domain accuracy, current state-of-the-art end-to-end detectors primarily memorize attack-specific statistical patterns seen during training. When confronted with evolving synthesis methods, particularly diffusion-based and prompt-conditioned systems that introduce different artifacts, detector performance collapses. Manual inspection further indicates that out-of-domain failures concentrate in two scenarios:

\begin{enumerate}
    \item High-fidelity diffusion TTS produces natural prosody, phase coherence, and breathing dynamics that differ fundamentally from GAN- and flow-based architectures, exposing features the detector fails to learn.
    \item Reliance on vocoder-based training systems biases the detector toward narrow spectral cues (e.g., phase discontinuities and formant distortions), hindering generalization to unseen synthesis methods.
\end{enumerate}

This generalization problem poses serious real-world risks. Adversaries can bypass detectors using synthesis methods absent from public training corpora, exploiting the rapid pace of TTS innovation. New architectures emerge monthly, while retraining cycles take months or years. Attackers may also combine multiple synthesis stages to produce hybrid artifacts unseen during training, an issue that incremental dataset expansion cannot solve. Since current detectors focus on attack-specific rather than general synthesis patterns, addressing this challenge requires not only continuously updated corpora, but, most importantly, model architectures that capture invariant synthesis characteristics.

\subsubsection{Cross-Lingual Evaluation}

To evaluate cross-lingual generalization beyond the training language (Chinese), we test the detection model on two large-scale English datasets, \textbf{ASVspoof 2021 LA and DF tracks} \cite{yamagishi2021asvspoof}, without retraining the detector. Table~\ref{tab:ood_detailed} shows modest performance degradation, demonstrating that the XLS-R multilingual frontend enables reasonable cross-lingual transfer from Mandarin-trained models to English deepfakes and provides a degree of robustness in detection.

\subsubsection{Robustness under Environmental Noise} 
Finally, we evaluate performance under environmental noise. As shown in Table~\ref{tab:noise}, detection performance drops sharply at a Signal-to-Noise Ratio (SNR) of 10 dB, with EER increasing from 0.83\% to 16.24\%, highlighting the model’s high sensitivity to noise. Using RawBoost, which augments training data with realistic environmental noise, the noisy EER is reduced to 2.55\% (6.4$\times$ improvement), and training on the pooled clean + noisy dataset achieves 1.53\% EER. These results confirm that noise-aware training is essential for robust real-world deployment.

\begin{table}[t]
\centering
\small
\begin{tabular}{lccc}
\toprule
Training & Clean & SNR=10 dB & Pooled \\
\midrule
Clean Data only      & 0.83 & 16.24 & 8.54 \\
\quad \textbf{+ RawBoost}      & \textbf{0.53} &  \textbf{2.55} & \textbf{1.53} \\
\bottomrule
\end{tabular}
\caption{Noise robustness with data augmentation.}
\label{tab:noise}
\end{table}

\section{Conclusion}
This paper presents a systematic evaluation of audio-based biometric authentication against contemporary voice cloning attacks, revealing, unfortunately, negative results that expose critical vulnerabilities in current defenses. The speaker verification system can be easily bypassed by open-source deepfake models trained on very small datasets, and the deepfake detection model, despite strong in-domain performance, fails to generalize to unseen attacks, highlighting risks from overestimated system security. Given that authentication systems protect millions of users across high-stakes domains, the rapid evolution of voice synthesis demands a fundamental shift in how we approach audio security. Future research should move beyond signature-based detection to learn invariant properties of synthesis and recognize that voiceprint authentication alone is unlikely to provide reliable protection. Defense-in-depth strategies that combine robust detection, multi-factor authentication, and adaptive measures stand as crucial safeguards against the next generation of audio deepfakes.

\section{Limitations}
Despite the comprehensive evaluation presented in this work, there are several limitations that offer opportunities for further research:

\begin{itemize}
    \item \textbf{Training data scale:} Our evaluation focuses on deepfake models trained with a small amount of target speaker data. While the results already provide a strong and urgent warning about security risks, studying how these risks scale with larger amounts of available data could reveal further insights. This is especially relevant in practice, as many individuals have substantial amounts of speech publicly accessible on the internet, potentially enabling even more effective attacks.

    \item \textbf{Complexity of cross-lingual evaluation:} Our study demonstrates cross-lingual transfer from Chinese to English, two languages that differ fundamentally in phonetic composition, tonal structure, and syllable patterns. While these experiments highlight the promise of multilingual self-supervised representations, evaluating additional languages with diverse phonological and prosodic characteristics, as well as the consideration of code-switching speech, could reveal further insights into generalization and robustness, guiding the design of truly language-agnostic defenses.
\end{itemize}

\bibliography{references}

\appendix

\section{Implementation Details}
\label{appendix:implementation}

The detection system is trained for 100 epochs with a batch size of 32 using AAM-softmax loss to enhance inter-class separability and intra-class compactness, optimized with AdamW (initial learning rate $3 \times 10^{-4}$) and a cosine annealing schedule to stabilize convergence. Input utterances are 4 seconds at 16 kHz, augmented with SpecAugment (time masking up to 40 frames, frequency masking up to 4 bands) and RawBoost (ISD SNR 10--40 dB, LnL gain 0--30 dB, SSI SNR 0--40 dB) to improve robustness under diverse acoustic conditions. Gradient clipping at norm 5.0 prevents instability. Training on 4× V100 GPUs takes approximately 10 hours, and inference runs at approximately 3 seconds per minute of audio on a single V100. The trained model and code will be publicly released to support efficient and reproducible evaluation.

To ensure robust implementation and reflect practical usage, we tested the detector across multiple audio formats and compression levels (Table~\ref{tab:codec}). Results show minimal performance variance (EER 0.83 -- 0.94\%), indicating that RawBoost augmentation effectively mitigates the impact of compression artifacts. This demonstrates that the system closely reflects a deployable real-world setup, ensuring reliable detection in scenarios such as phone banking, where audio may pass through multiple codec stages.

\begin{table}[t]
\centering
\begin{tabular}{lc}
\toprule
Format & EER (\%) \\
\midrule
Raw (16 kHz)         & 0.83 \\
WAV (uncompressed)   & 0.91 \\
MP3 (245 kbps)       & 0.84 \\
MP3 (100 kbps)       & 0.94 \\
OGG (160 kbps)       & 0.84 \\
OGG (256 kbps)       & 0.86 \\
\bottomrule
\end{tabular}
\caption{Codec compression robustness.}
\label{tab:codec}
\end{table}

\section{Baseline Deepfake Detection Models}
\label{appendix:baselines}

In Table \ref{tab:baselines}, we compare a set of representative baseline deepfake detection models spanning the evolution of audio anti-spoofing techniques, alongside the state-of-the-art XLS-R + AASIST architecture, to provide a comprehensive overview of system performance. Below, we provide further details and justifications for the baseline selection.

Linear Frequency Cepstral Coefficients with Gaussian Mixture Models (LFCC + GMM) provide a classical statistical baseline using hand-crafted features and generative modeling, widely employed in early ASVspoof challenges \cite{todisco2018integrated}. ResNet34 with spectrogram input serves as a standard CNN-based baseline, leveraging residual learning to capture discriminative time-frequency patterns \cite{he2016deep}. RawNet2 is an end-to-end model operating on raw waveforms with learnable filterbanks, demonstrating effective data-driven feature learning without explicit feature extraction \cite{tak2021end}.

Audio Anti-Spoofing using Integrated Spectro-Temporal Graph Attention Networks (AASIST) extends RawNet2 by jointly modeling spectro-temporal spoofing artifacts via graph attention, representing a state-of-the-art standalone countermeasure without self-supervised pretraining \cite{jung2022aasist}. When combined with XLS-R, the system incorporates multilingual self-supervised representations, achieving strong performance on the in-domain test set. However, it still suffers from limited generalization to unseen attacking models, a critical vulnerability that poses significant security risks in practical deployment.

\section{Deepfake Model}
\label{appendix:model list}

To expand the benchmark test set, we include speech generated from the following TTS systems:

\begin{itemize}
    \item \textbf{AdaSpeech4}~\cite{wu2022adaspeech}: A zero-shot adaptive TTS system that synthesizes speech for unseen speakers using factorized speaker representations, achieving high naturalness and similarity.
    
    \item \textbf{BinauralGrad}~\cite{leng2022binauralgrad}: A two-stage conditional diffusion model for binaural audio synthesis, capturing spatial cues for realistic spatialized sound.
    
    \item \textbf{MPBert}~\cite{zhang2022mixed}: Enhances TTS by integrating mixed phoneme and sub-phoneme embeddings, improving phonetic detail and pronunciation accuracy.
    
    \item \textbf{NaturalSpeech 1}~\cite{tan2024naturalspeech}: End-to-end TTS model achieving human-level quality and expressivity by jointly modeling the entire synthesis pipeline.
    
    \item \textbf{NaturalSpeech 2}~\cite{shennaturalspeech}: Utilizes neural audio codec and latent diffusion for natural speech and singing, supporting zero-shot generation for new speakers.
    
    \item \textbf{NaturalSpeech 3}~\cite{ju2024naturalspeech}: Factorizes speech into content, prosody, timbre, and acoustic subspaces with a neural codec and diffusion model, improving zero-shot quality and speaker similarity.
    
    \item \textbf{PromptTTS 1}~\cite{guo2023prompttts}: Generates speech from natural language prompts describing content and style, enabling controllable and flexible TTS.
    
    \item \textbf{PromptTTS 2}~\cite{lengprompttts}: Builds on PromptTTS with prompt-driven style modeling and enhanced variability, producing consistent and expressive synthetic voices.
\end{itemize}

These models collectively span diverse synthesis paradigms, including adaptive, diffusion-based, and prompt-conditioned approaches,  ensuring that out-of-domain evaluation captures realistic and challenging variations in modern speech synthesis while revealing true security robustness.

\end{document}